# Performance Comparisions of ICA Algorithms to DS-CDMA Detection

Sargam Parmar, Bhuvan Unhelkar

**Abstract**— Commercial cellular networks, like the systems based on DS-CDMA, face many types of interferences such as multi-user interference inside each sector in a cell to interoperate interference. Independent Component Analysis (ICA) has been used as an advanced preprocessing tool for blind suppression of interfering signals in DS-CDMA communication systems. The role of ICA is to provide an interference-mitigated signal to the conventional detection. This paper evaluates the performance of some major ICA algorithms like Cardoso's joint approximate diagonalization of eigen matrices (JADE), Hyvarinen's fixed point algorithm and Comon's algorithm to solve the symbol estimation problem of the multi users in a DS-CDMA communication system. The main focus is on blind separation of convolved CDMA mixture and the improvement of the downlink symbol estimation. The results of numerical experiment are compared with those obtained by the Single User Detection (SUD) receiver, ICA detector and combined SUD-ICA detector.

———————————————— ◆ ————————————————

## 1 INTRODUCTION

THIS paper evaluates the performance of some major ICA algorithms to solve the symbol estimation problem of the multi users in a DS-CDMA communication system. The need to solve this problem arises because wireless communication networks and systems such as those used in mobile phones, have an essential challenge in division of this common transmission medium among several users. A primary goal of communication system is to enable each user of the system to communicate reliably despite the fact that the other users share the same resources, possibly simultaneously. As the number of users in the system grows, it becomes necessary to improve the efficiency of these common communication resources.

Various communication systems based on CDMA (Code Division Multiple Access) techniques have become popular, because they offer several advantages over the more traditional FDMA and TDMA schemes based on the use of non-overlapping frequency or time slots assigned to each user. The capacity of CDMA based communication system is larger, and it degrades gradually with increasing number of simultaneous users who can be asynchronous. CDMA systems require more advanced signal processing methods, and correct reception of CDMA signals is more difficult because of several disturbing phenomena such as multipath propagation, possibly fading channels, various types of interferences, time delays, and different powers of users [1],[2],[3].

In a Direct Sequence-Code Division Multiple Access (DS-CDMA) system, users share the same band of frequencies and the same time slots, but they are separated in code. In downlink signal processing, each mobile station knows only its own code while the codes of the others stations are unknown. There is also less processing power in the mobile station than in the base station. These features of downlink processing call for new, efficient and simple solutions. Independent Component Analysis (ICA) is a recently developed, useful extension of standard Principal Component Analysis (PCA)[4]. ICA has been proposed and developed to take advantage of Blind Source Separation (BSS) algorithm. ICA and BSS techniques provide a promising new approach to the downlink signal processing of DS-CDMA systems using short spreading codes [4][5].

Jutten and H´erault provided one of the first significant approaches to the problem of blind separation of instantaneous linear mixtures [6]. Since then, many different approaches have been attempted by numerous researches using neural networks, artificial learning, higher order statistics, minimum mutual information, beam-forming and adaptive noise cancellation, each claiming various degrees of success. In this paper, our goal is a direct symbol estimation of the DS-CDMA downlink signal by means of Comon's[7], JADE [9], and FastICA [11] based methods. The results of numerical experiment are compared with those obtained by the conventional match filter or Single User Detection (SUD) receiver.

## 2 MATERIAL AND METHODS

We consider the classical ICA model with instantaneous mixing

$$x = As \qquad (2.1)$$

where the sources $s = [s1, s2, ..., sn]^T$ are mutually independent random variables and $A_{n \times n}$ is an unknown invertible mixing matrix. The goal is to find only from observations, x, a matrix W such that the output

$$y = Wx \qquad (2.2)$$

is an estimate of the possible scaled and permutated source

• F.A. Author is with the U.V.Patel College of Engineering, Ganpat University, Kherva-382711,India.
• S.B. Author is with the school of Computing and Mathematics, University of Western Sydney,Australia.

vector s.

Several algorithms exits for blind source separation. We evaluate the performance of JADE, Comon's and FastICA algorithm to DS-CDMA communication system. This section presents a brief description of the respective approaches of the ICA algorithms.

## A. Comon's Algorithm

A specific contrast function is proposed, based on minimization of mutual information between the components at the output of separator (which is directly related to Kullback-Leibler divergence between the output vector probability density function (pdf) and it's pdf if it was made of independent components)[7][8]. After some manipulations on the Edgeworth expansion of the source joint pdf, the contrast function simplifies into sum of the output squared $r$th-order marginal cumulants and for $r = 4$ it becomes output squared kurtosis:

$$\Psi_4(\mathbf{Q}) = \sum_{i=1}^{q} (k s_{iiii})^2 \qquad (2.3)$$

## B. JADE Algorithm

The JADE algorithm [9][10] relies on second and fourth-order cumulants to separate the sources. SOS is used to obtain a whitening matrix $\mathbf{Z}$ from the sample covariances. To reduce the computational load, only the $n$ most significant eigen pairs of fourth order cumulants obtained from the whitened process are joint diagonalized by unitary matrix $\mathbf{U}$. The separated matrix can be estimated as $\mathbf{U}^{\dagger}\ \mathbf{Z}$, where $\dagger$ represents pseudo inverse. The JADE contrast function is the sum of squared fourth order cross cumulants

$$\Phi^{JADE(Y)} = \sum_{ijkl \neq iikl} (\mathbf{Q}_{ijlk})^2 \qquad (2.4)$$

As this algorithm uses cross-cumulants, there is no need to go for gradient descent and hence there is no chance of divergence.

## C. Fixed-Point Algorithm

The original fixed-point algorithm [11] uses kurtosis and computations can be performed either in batch mode or in a semi-adaptive manner. It uses deflation approach to update the columns of separating matrix $\mathbf{W}$ and to find the independent components one at a time. More recent versions are using hyperbolic tangent, exponential or cubic functions as contrast function. The update rule for the deflation method is given by [12]

$$\mathbf{w}^*(k) = \mathbf{C}^{-1} E\{\mathbf{x}\ g(\mathbf{w}(k-1)^T\ \mathbf{x})\} - E\{g'(\mathbf{w}(k-1))^T\ \mathbf{x}\}\ \mathbf{w}(k-1) \qquad (2.5)$$
$$\mathbf{w}(k) = \mathbf{w}^*(k)/\ ?\mathbf{w}^*(k)^T \mathbf{C}\ \mathbf{w}^*(k) \qquad (2.6)$$

where $g$ can be any suitable non-quadratic contrast function, with derivative $g'$; and $\mathbf{C}$ is the covariance matrix of the mixtures, $\mathbf{x}$. $\mathbf{w}(k)^T\ \mathbf{x}(t)$; $t = 1, 2, \ldots$ equals one of the sources.

## D. System Model

The signal model of a downlink DS-CDMA receiver over multipath fading channel has the form

$$\mathbf{R} = \mathbf{GB} + \mathbf{N} \qquad (2.7)$$

Where, matrices G, $\mathbf{B}$ and $\mathbf{N}$ denotes the unknown mixed matrix, symbols and noise.

$$R = [r_m, \ldots, r_{m'}]$$

$$B = [b_m, \ldots, b_{m'}]$$

$$N = [n_m, \ldots, n_{m'}]$$

where R, N is CxM' matrix, G is Cx2KL matrix and B is 2KLxM' matrix.

Comparing with the model of linear ICA in Eq. 2.1, $\mathbf{B}$ is the source signal $\mathbf{s}$ need to be estimated, $\mathbf{R}$ is the observed mixed signal $\mathbf{x}$, and G is the unknown mixing matrix A. The noise matrix $\mathbf{N}$ in Eq. 2.7 can be treated as an independent component to be added into $\mathbf{x}$.

In this paper, the code timing and channel estimation are assumed the prerequisite tasks. In other words, the received signal for ICA detector is assumed the sampled and synchronized data. Thus the path is down to 1; the propagation delay is equal to 0. In view of ICA algorithm, the AWGN can be treated as one of the independent components. Then the Eq. 2.7 is changed to

$$\mathbf{R} = \mathbf{GB} \qquad (2.8)$$

where the dimension of matrices $\mathbf{R}$, $\mathbf{B}$ and G are CxM', CxK and K xM', respectively.

## 3 NUMERAICAL EXPERIMENT

The algorithms were tested using simulated DS-CDMA downlink data in the presence of AWGN or Pink Noise. The short Gold code to make the length of chips to be C = 31, was used. Thus the maximum number of users is K = 30 as the 31st user is AWGN or Pink Noise. Monte Carlo simulation that has incorporated the FastICA, JADE and Comon's algorithm were run to verify the validity of the system model.

A combined scheme of the ICA-SUD detector proposed [4] was devised and implemented. This combined scheme is illustrated in Fig 1. One independent ICA detectors are incorporated in parallel with a SUD detector. The majority of the one ICA and one SUD detectors make final decisions on the outcomes. The results of numerical experiment are shown in Fig. 2-11. The performance appears to be of average symbol-error-rate (SER).

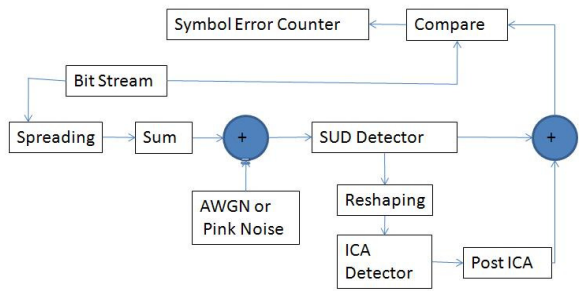

Figure 1 SUD-ICA Detector

For AWGN Parameters were set as follows: In order to test the performance of the ICA detector, the values of Number of symbol M were varied for M = 2000, 5000 and 10000; Number of users K = 30; Number of paths L = 1. Signal-to-Noise Ratio (SNR) was varied with respect to the individual user from -10dB to 0dB. All the signals for each of users are sent in same power. The AWGN is treated as an independent component (IC). One of the typical simulation results of the ICA detector is shown in Fig. 2. 3. 4 along with the results of the SUD receiver, ICA receiver and combined SUD and ICA- SUDICA receiver.

We can easily observe that the ICA detector is functionally working. It can semi-blindly estimate DS-CDMA downlink signal. However, when the decoding is blind, it requires training sequences to match the decoded bit stream to the user who should receive it. However, in this simulation, more than 10% of simulation runs failed to converge, because some of IC's were not found during the iterations. ICA receiver alone can not work well for presence of Gaussian noise. A combined scheme of the ICA detector, SUD detector and SUDICA detector was devised and implemented. The results of numerical experiment of the SUD, ICA, and SUDICA detectors are shown in Fig. 2. 3. 4. The performance appears to be of average bit-error-rate (BER). However, the ICA-SUD detector outperforms the ICA detector giving the lowest BER to the SUDICA detector as shown in Fig. 2. 3. 4. Fig. 5. 6. Show comparision of ICA detectors and SUDICA detectors for JADE, FastICA and Comon's algorithm. JADE algorithm performs well compare to FastICA and Comon's, which gives BER as same as SUD detector.

For Pink Noise Parameters were set as follows: the Number of symbol M were varied for M = 2000, 5000 and 10000; Number of users K = 30; Number of paths L = 1. Signal-to-Noise Ratio (SNR) was varied with respect to the individual user from -10dB to 0dB. All the signals for each of users are sent in same power. The Pink Noise is treated as an IC. One of the typical simulation results of the ICA detector is shown in Fig. 7. 8. 9 along with the results of the SUD receiver, ICA receiver and combined SUD and SUDICA receiver.

We can easily observe that the ICA detector is functionally working. It can semi-blindly estimate DS-CDMA downlink signal. However, when the decoding is blind, it requires training sequences to match the decoded bit stream to the user who should receive it. However, in this simula-tion, more than 10% of simulation runs failed to converge, because some of IC's were not found during the iterations. ICA receiver alone can work well for presence of Pink noise compared to SUD or SUDICA detector for FastICA and JADE but not for Comon. The results of numerical experiment of the ICA, SUDICA detector are shown in Fig. 10. 11 for comparison. The performance appears to be of average bit-error-rate (BER). SUDICA detector is able to improve the performance of the SUD method only marginally. SUDICA is better than SUD for SNR=-10dB and -5dB for FastICA and JADE but not for Comon.

The number of symbols M involved in a single run of ICA detection is usually M = 1000 as in the cases of literatures [13][14][4]. In order to test the performance of the ICA detector, the values of M were varied for M = 2000, 5000 and 10000. Simulation results are summarized in Fig. 2-12. We can observe that the simulation results have not more effect on SER. The computational load gets heavier as M increases, since about 100 simulations are performed for each value of M.

# 4 CONCLUSIONS

In this paper we have evaluated the performance of some major ICA algorithms like FastICA, Cardoso's joint approximate diagonalization of eigen matrices (JADE), and Comon's algorithm to solve the symbol estimation problem of the multi users in a DS-CDMA communication system. We observe that the performances of the algorithms are affected by noise color. The performance of the FastICA and JADE algorithm are slightly better compared to the Comon algorithm. ICA based DS-CDMA downlink detector demonstrated that ICA detector can solve the symbol estimation problem with no spreading code required, though the spreading code should be utilized to identify each user. Thus an ICA, SUDICA detector have been used and their symbol error rate is lower than the conventional SUD detector concluded from numerical experiments. Even if the powers of the signals are the same, additional multiple access interference can be mitigated by ICA, thus improving the performance of SUD. When the number of symbols M involved in a single simulation run is increased from 2000 to 10000, the number of iteration is significantly reduced from 20 down to 5 in average. With the increase in the number of symbols, the symbol error rate is not much affected. As well, the increased amount of computational time by a larger M can be compensated by a less number of iterations required.

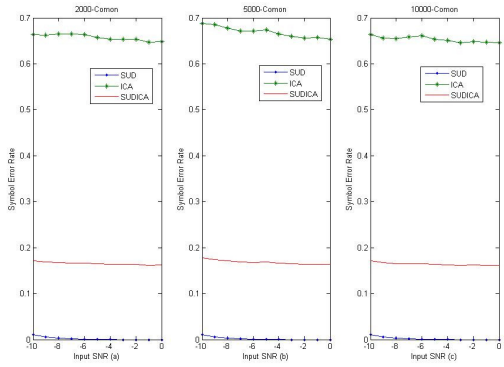

Figure 2 Output SER in presence of Gaussian noise using Comon's algorithm for number of symbols (a) 2000 (b) 5000 (c) 10000

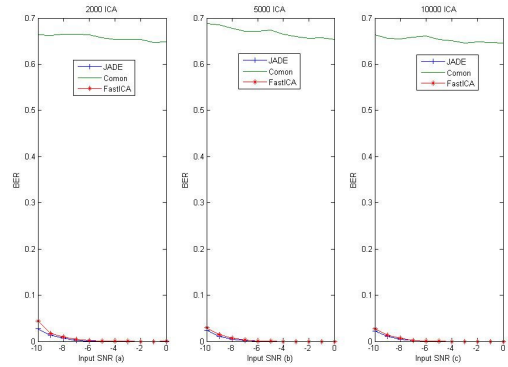

Figure 5 Comparisons of ICA Detecotr in presence of Gaussian noise using ICA algorithms for number of symbols (a) 2000 (b) 5000 (c) 10000

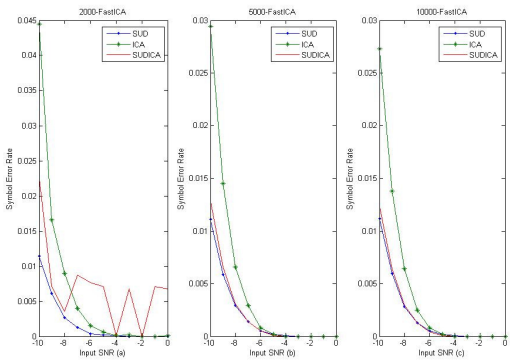

Figure 3 Output SER in presence of Gaussian noise using FastICA algorithm for number of symbols (a) 2000 (b) 5000 (c) 10000

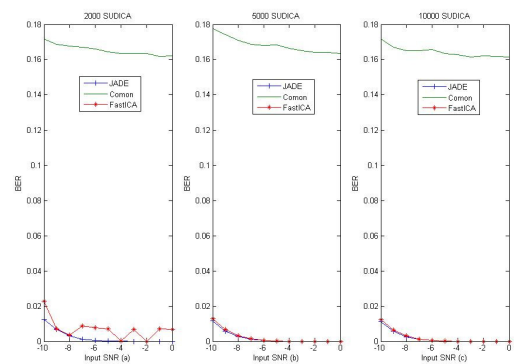

Figure 6 Comparisons of SUD-ICA Detector in presence of Gaussian noise using ICA algorithms for number of symbols (a) 2000 (b) 5000 (c) 10000

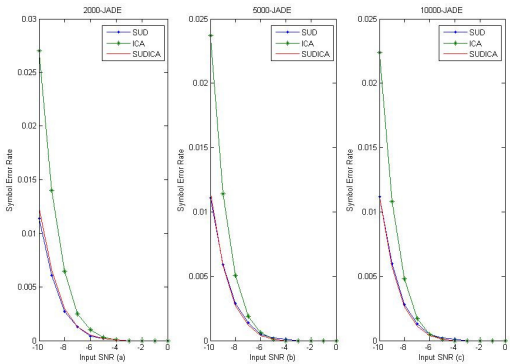

Figure 4 Output SER in presence of Gaussian noise using JADE algorithm for number of symbols (a) 2000 (b) 5000 (c) 10000

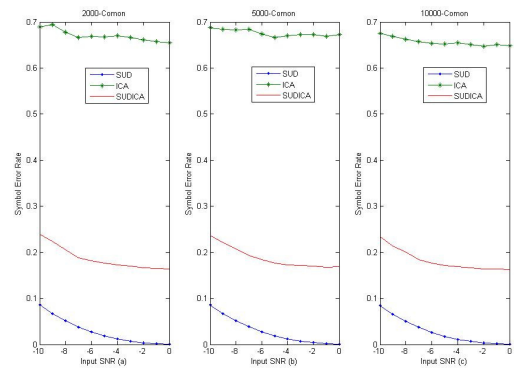

Figure 7 Output SER in presence of Pink noise using Comon's algorithm for number of symbols (a) 2000 (b) 5000 (c) 10000

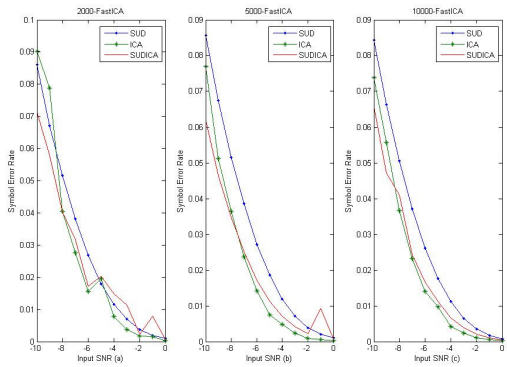

Figure 8 Output SER in presence of Pink noise using FastICA algorithm for number of symbols (a) 2000 (b) 5000 (c) 10000

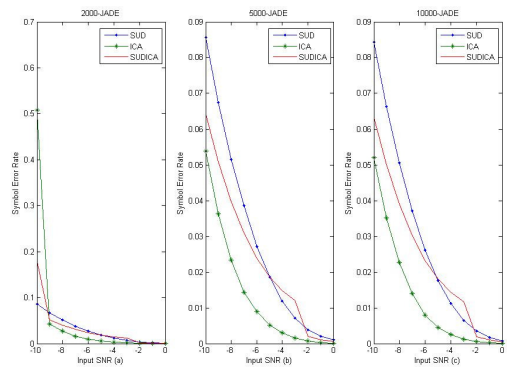

Figure 9 Output SER in presence of Pink noise using JADE algorithm for number of symbols (a) 2000 (b) 5000 (c) 10000

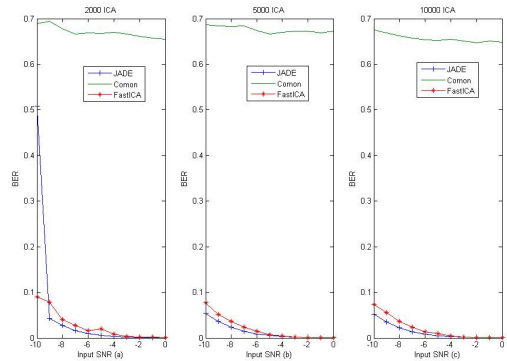

Figure 10 Comparisons of ICA Detector in presence of Pink noise using ICA algorithms for number of symbols (a) 2000 (b) 5000 (c) 10000

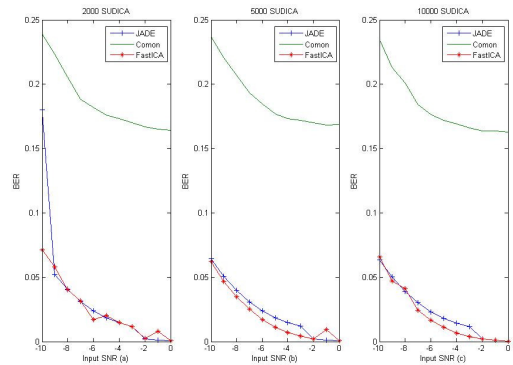

Figure 11 Comparisons of SUDICA Detector in presence of Pink noise using ICA algorithms for number of symbols (a) 2000 (b) 5000 (c) 10000